\documentclass[11pt]{article}
\usepackage{amsmath,amssymb,amscd}
\usepackage{amsthm,bbm}
\usepackage{mathtools}
\usepackage{float}
\usepackage[caption = false]{subfig}
\usepackage[final]{graphicx}
\usepackage[mathscr]{eucal}
\usepackage{color}
\usepackage{booktabs}
\newtheorem{Proposition}{Proposition}
\large\normalsize

\title{Stochastic viability in an island model with partial dispersal : Approximation by a diffusion process in the limit of a large number of islands}
\author{Dhaker Kroumi$^1$\footnote{Author for
correspondence, and e-mail: dhaker.kroumi@kfupm.edu.sa} and Sabin Lessard$^2$
\\$^1$Department of Mathematics and Statistics\\King Fahd University of Petroleum and Minerals\\Dhahran 31261, Saudi Arabia\\
$^2$Department of Mathematics and Statistics\\University of Montreal \\
 Montreal H3C 3J7, Canada\\
 }
\date{}
\begin{document}
\maketitle


\section*{Abstract}

In this paper, we study a finite population undergoing discrete, nonoverlapping generations, that is structured into $D$ demes, each containing $N$ individuals of two possible types, $A$ and  $B$, whose viability coefficients, $s_A$ and $s_B$, respectively, vary randomly from one generation to the next. We assume that the means, variances and covariance of the viability coefficients are inversely proportional to the number of demes $D$, while higher-order moments are negligible in comparison to $1/D$.
We use a discrete-time Markov chain with two time scales to model the evolutionary process, and we demonstrate that as the number of demes $D$ approaches infinity, the accelerated Markov chain converges to a diffusion process for any deme size $N\geq 2$. This diffusion process allows us to evaluate the fixation probability of type $A$ following its introduction as a single mutant in a population that was fixed for type $B$.
We explore the impact of increasing the variability in the viability coefficients on this fixation probability. At least when $N$ is large enough, it is shown that increasing this variability for type $B$ or decreasing it for type $A$ leads to an increase in the fixation probability of a single $A$. The effect of the population-scaled variances,  $\sigma^2_A$ and $\sigma^2_B$, can even cancel the effects of the  population-scaled means, $\mu_A$ and $\mu_B$.
We also show that the fixation probability of a single $A$ increases as the deme-scaled migration rate increases.  Moreover, this probability is higher for type $A$ than for type $B$ if the population-scaled geometric mean is higher for type $A$ than for type $B$, which means that $\mu_A-\sigma_A^2/2>\mu_B-\sigma_B^2/2$.


\noindent \textbf{Keywords and phrases}: Structured population; Fixation probability; Variability in selection coefficients; Island Model; Diffusion approximation. 

\noindent \textbf{Mathematics Subject Classification (2010)}: Primary 92D25; Secondary 60J70


\section{Introduction}
Initial studies in population genetics focused on well-mixed populations and typically consider two reproduction schemes: the Moran model described in Moran \cite{M1958}, and the Wright-Fisher model first proposed by Fisher \cite{F1930} and Wright \cite{W1931}.

Under the Moran model, at each time step, a parent is selected with probability proportional to its fitness  to produce an offspring, and this offspring replaces a randomly chosen individual from the population. This process can be analyzed using a birth-death process, even for small populations, and can yield valuable genetic insights on the population state or its evolution, such as the fixation probability for a given type  in the absence of mutation or its average frequency in the stationary state in the presence of recurrent mutation. 
 
 In contrast, under the Wright-Fisher model, each individual produces a large number of offspring proportional to its fitness, and a fixed number of individuals are randomly sampled to form the next generation. While this process cannot be readily analyzed for any given population size, an approximation for large population using a diffusion process (Ito and Mckean \cite{IM1965}) enables the calculation of useful biological quantities such as fixation probabilities and times to fixation.

Diffusion processes are a class of mathematical models commonly used to describe the behaviour of genetic systems (see chapter 15 in Karlin and Taylor \cite{KT1981}). These models aim to capture the interactions between multiple genes and their respective biological processes over time. In particular, these models are used to approximate the behaviour of large genetic systems, where the interactions between genes are too complex to be described directly. Note that for large populations, the Moran model with appropriate time scale and selection intensity can be approximated by the same Wright-Fisher diffusion process, rendering the two models equivalent. 

These models oversimplify the reality of biological populations, which are more complex. Evidence of geographic structure in genetic data, as demonstrated by Slatkin \cite{S1985}, highlights the importance of 
studying populations partitioned into different groups subject to specific environmental and ecological effects and connected through gene flow.

One of the basic models for spatially structured populations is the island model, which assumes a population that is subdivided into many isolated demes that are exchanging migrants. Ethier and Nagylaki \cite{EN1980} established key conditions that must be met by a discrete-time Markov chain with two time scales in order to infer its weak convergence to a diffusion process in the limit of a large population. This approximation is crucial for studying the effects of migration and selection on genetic diversity and its evolution in geographically structured natural populations from a theoretical perspective. Nagylaki \cite{N1980} explored these effects on populations that reproduce according to the Wright-Fisher model with random mating. Specifically, he focused on a fixed number of demes that exchange migrants according to a constant migration matrix that satisfies ergodicity conditions. When the mutation rates and selection intensities are inversely proportional to the population size, a diffusion approximation can be validated as the size of each deme tends towards infinity. This approximation corresponds to a strong migration limit. 
Later on, Nagylaki \cite{N1989} used a diffusion equation to describe the changes in allele frequencies over time due to migration and selection, where he showed that the effect of selection can be either enhanced or diminished by migration depending on the strength of the two forces. Nagylaki \cite{N1996} extended his diffusion model for migration and selection to take into account the effects of dioecy, a reproductive system in which individuals are either male or female. This requires a set of equations to describe the changes in allele frequencies due to migration, selection, and sex-specific differences in reproductive success. He showed that dioecy can have a significant impact on the genetic structure of populations, particularly when there are sex-specific differences in selection pressures or migration rates. Moreover, Nagylaki \cite{N1997} applied his diffusion model to a plant population to describe the changes in allele frequencies due to migration, selection, and factors such as self-fertilization and inbreeding.

Based on numerical simulations, Cherry \cite{C2003a,C2003b} and Cherry and Wakeley \cite{CW2003} showed the applicability of diffusion methods in studying the impact of dominance, population structure, and local extinctions on fixation probabilities of mutant alleles in island models. Whitlock \cite{W2003} applied  similar methods to the stepping-stone model of population structure. The stepping-stone model assumes that the population is subdivided into a series of demes, where each deme has a fixed size and is connected to its neighbouring demes by migration. 

Roze and Rousset \cite{RR2003} proposed a method for constructing diffusion approximations in structured populations. This method uses general expressions for the expectation and variance in allele frequency change over one generation in terms of partial derivatives of the fitness function and probabilities of genetic identity under a neutral model. They used this method to derive the fixation probability of new mutant alleles based on their dominance coefficient, the partial selfing rate, and deme extinction rate.

Wakeley \cite{W2003} and Wakeley and Takahashi \cite{WT2004} studied an island model with a finite number of demes of the same finite size in the limit of a large number of demes. In a haploid population, individuals are categorized as either of type $A$ or of type $B$, and each of the $D\geq2$ demes has $N\geq2$ individuals. Viability selection for each type is inversely proportional to $D$. Wakeley \cite{W2003} proposed a Wright-Fisher reproduction scheme within each deme with a migration pool where each deme contributes equally. Then, a portion $m>0$ of each deme is replaced by offspring coming from the migration pool. This corresponds to a soft selection scenario (Christiansen \cite{C1975}).
Wakeley and Takahashi \cite{WT2004} studied a Moran reproduction scheme, where at each time step a random deme is chosen from which an individual is selected to die proportionally to its relative death rate. Then, another individual is chosen to produce a copy of itself which will take the vacant position. This parent is chosen from the same deme as the individual selected to die with probability $1-m$, or from another deme with probability $m$. In both models, the frequency of type $A$ in the population can be approximated through a diffusion process, as long as key conditions described in Ethier and Nagylaki's \cite{EN1980} are fulfilled. 

The approximation in Wakeley \cite{W2003} depends on a conjecture regarding the equilibrium state in the case of an infinite number of demes 
in the absence of selection where the average frequency of type $A$ in the population, $x$, remains constant over time.
Let $(v_i)_i$ be the probability distribution of such an equilibrium, where $v_i$ is the frequency of demes of type $i$ in a population where the frequency of $A$ is $x$. Then, in one time step, the new frequency of demes of type $i$ can be written in the form
\begin{equation}
v_i(1)=v_i+\sum_{k=1}^Nr_k\left(X(1)-x\right)^{k},
\end{equation}
where $X(1)$ is the new frequency of type $A$ and $r_k$ for $k=1, \ldots, N$ depend only on $N$, $m$ and $x$. Lessard \cite{L2007} confirmed the validity of this conjecture by using the Ewens sampling formula  for the infinitely-many-alleles model with an analogy argument between mutation events and migration events. Lessard \cite{L2009} extended Wakeley model \cite{W2003} to a diploid population with multiple types, considering mutation and various scenarios of migration and selection.

All the models discussed above assume constant viability coefficients, and this supposes that the environment remains unchanged over time. However, this is unrealistic because environmental conditions can fluctuate randomly, leading to changes in competition capabilities and birth-death rates (Kaplan \textit{et al.} \cite{KHH1990}, Lande \textit{et al.} \cite{LES2003}, May \cite{M1973}).
These fluctuations can affect the population size and composition over time, and many researchers have investigated their impact. For instance, Lambert \cite{L2006}, Parsons and Quince \cite{PQ2007a,PQ2007b}, and Otto and Whitlock \cite{OW1997} have examined the fixation probability for a mutant type in an unstructured population whose size fluctuates dynamically due to various demographic scenarios of growth or decline. On the other hand, Uecker and Hermisson \cite{UH2011} have focused on the case of a single beneficial allele in a population that undergoes temporal variation in its size and selection pressure.

Another feature worth considering is the possibility of variability in selection coefficients, which pertains to the fluctuation in fitness among different types. 
Numerous studies have investigated the impact of different selection coefficients between generations or offspring numbers within generations in both haploid and diploid population genetic models without structure. Gillespie \cite{G1973} proposed a mathematical model to study the variation in selection coefficients between generations in an infinite haploid population. Gillespie \cite{G1974} also examined how natural selection can favor variability in the number of offspring produced within a generation, considering both genetic and environmental sources of variation, and showed that selection can favor individuals that produce a mix of high and low numbers of offspring. Karlin and Levikson \cite{KLe1974} explored how selection pressures can vary over time in small populations and demonstrated that fluctuations in selection intensity can be significant and can lead to the fixation or loss of alleles that would not be affected by constant selection pressures. Karlin and Liberman \cite{KLi1974} examined the effects of random fluctuations in selection intensity on large populations. Starrfelt and Kokko \cite{SK2012} considered the case where organisms reduce their fitness variance at the expense of their mean fitness, to increase their chances of survival in unpredictable or fluctuating environments. They have shown that a trade-off exists between mean fitness, variance, and correlations among different fitness components, and that the optimal bet-hedging strategy depends on the nature of environmental fluctuations. Some extensions of these studies can be found in Schreiber \cite{S2015} and Rychtar and Taylor \cite{RT2019}.

Studying the effect of variability in selection viability coefficients in structured populations is crucial for understanding the dynamics of natural populations. In such populations, individuals are more likely to interact with individuals from their own subpopulation, leading to localized adaptation and potentially increasing genetic differentiation. Understanding the patterns and drivers of this variability can provide insights into the evolution and persistence of structured populations.

This paper will analyze the impact of such variability in a model of a population structured into a large number of isolated demes of the same finite size interconnected by migration. This model is the one presented in Wakeley \cite{W2003}, but with the inclusion of variability in selection coefficients. Specifically, we will investigate the combined effects of the second moments in selection intensities and the migration rate on the evolution of a particular type, particularly in its fixation probability under various scenarios.

The structure of this paper is as follows: Section 2 presents the model, while Section 3 provides the proof of the conditions for a diffusion approximation with two time scales. Appendices A and B contain some mathematical analysis related to the proof. In Section 4, we examine the fixation probability for the general case. Section 5 investigates the fixation probability under various scenarios when the size of each deme is large. Finally, we discuss our results and their relationship with the existing literature.

\section{Model}
Consider a population structured into $D$ demes, where each deme contains exactly $N$ individuals. Each individual can be either of the wild type, denoted by $B$, or of the mutant type, denoted by $A$.
A deme is said to be of type $i$ if it contains $i$ individuals of type $A$ and $N-i$ individuals of type $B$, for $i=0,1,\ldots,N$. We assume discrete, nonoverlapping generations and let $Z_i(t)$ be the fraction of demes of type $i$ at the beginning of generation $t\geq0$. Then, the population state is represented by the frequency vector $\mathbf{Z}(t)=(Z_0(t),Z_1(t),\ldots,Z_N(t))$, where $\sum_{i=0}^{N}Z_i(t)=1$. Moreover,
\begin{align}
X(t)= \sum_{i=0}^{N}\frac{i}{N}Z_i(t)
\end{align}
gives the corresponding frequency of type $A$ in the population.

The individuals in the different demes at the beginning of generation $t+1$ are obtained from the individuals in the previous generation as follows. First, every individual at the beginning of generation $t$ 
produces the same very large number of offspring. 
Then, a fraction $m$ of offspring uniformly disperse among all the demes, while a proportion $1-m$ stay in the deme where they were produced. This kind of dispersal is called \emph{proportional dispersal} (see, e.g., Lessard \cite{L2009}). After dispersal, there is viability selection within demes. It is assumed that the survival probability of 
an offspring depends only on its type, being proportional to $1+s_{A}$ if the offspring is of type $A$, or $1+s_{B}$ if the offspring is of type $B$. Moreover, it is assumed that the viability coefficients $s_{A}$ and $s_{B}$ are random variables, as a result of stochastic fluctuations from one generation to the next, whose moments are expressed as
\begin{subequations}\label{sec1-eq1}
\begin{align}
E\left[s_{A}\right]&=\frac{\mu_{A}}{ND}+o\left(D^{-1}\right),\\
E\left[s_{B}\right]&=\frac{\mu_{B}}{ND}+o\left(D^{-1}\right),\\
E\left[s_{A}^2\right]&=\frac{\sigma^2_{A}}{ND}+o\left(D^{-1}\right),\\
E\left[s_{B}^2\right]&=\frac{\sigma^2_{B}}{ND}+o\left(D^{-1}\right),\\
E\left[s_{A}s_{B}\right]&=\frac{\sigma_{AB}}{ND}+o\left(D^{-1}\right).
\end{align}
\end{subequations}
As for their higher-order moments,  we have
\begin{equation}\label{sec1-eq2}
E\left[|s_{A}|^{j}|s_{B}|^{k}\right]=o\left(D^{-1}\right),
\end{equation}
for any nonnegative integers $j$ and $k$ such that $j+k\geq 3$. For similar assumptions in unstructured populations, see Gillespie \cite{G1973,G1974}, Karlin and Levikson \cite{KLe1974}, Karlin and Liberman \cite{KLi1974} and Avery \cite{A1977} for population genetics models, and Li and Lessard \cite{LL2020} for evolutionary games. See also Soares and Lessard \cite{SL2019} for a model in an age-structured population. Here, the parameters $\mu_A$, $\mu_B$, $\sigma^2_A$, $\sigma^2_B$ and $\sigma_{AB}$ are population-scaled first and second moments in the limit of a large number of demes. Finally, the individuals to start generation $t+1$ in each deme are obtained by random sampling of $N$ offspring within the deme.

Suppose that the population state at the beginning of generation $0$ is given by the frequency vector $\mathbf{Z}^{}(0)=\mathbf{z}=(z_0,z_1,\ldots,z_N)$ and let 
\begin{align}
X^{}(0)=x=\sum_{i=0}^{N}\frac{i}{N}z_i
\end{align} 
be the corresponding frequency of type $A$ in the population. After reproduction and dispersal of offspring, the frequency of type $A$ in a deme of type $i$ is transformed into
\begin{equation}\label{sec1-eq3}
\tilde{x}_i=(1-m)\frac{i}{N}+mx,
\end{equation}
for $i=0,1,\ldots,N$. Then, viability selection with given coefficients $s_A$ and $s_B$ in generation $0$ will change this frequency into
\begin{align}
\tilde{\tilde{x}}_i&=\frac{\left(1+s_A\right)\tilde{x}_i}{\left(1+s_A\right)\tilde{x}_i+\left(1+s_B\right)(1-\tilde{x}_i)}=\frac{\left(1+s_A\right)\tilde{x}_i}{1+\tilde{s_{i}}},
\end{align}
where $\tilde{s}_i=\tilde{x}_is_{A}+(1-\tilde{x}_i)s_{B}$ is the average viability coefficient in the deme.

After random sampling of $N$ offspring, a deme that was of type $i$ at the beginning of generation $0$ will become a deme of type $j$ at the beginning of generation $1$ with conditional probability given by
\begin{equation}\label{sec1-eq4}
P_{ij}(\mathbf{z})=\binom{N}{j}\left(\tilde{\tilde{x}}_i\right)^j\left(1-\tilde{\tilde{x}}_i\right)^{N-j},
\end{equation}
for $i,j=0,1,\ldots,N$.

Let $Z_{ij}^{}(1)$ be the fraction of demes of type $j$ at the beginning of generation $1$ that were of type $i$ at the beginning of generation $0$ for $i, j =0, 1, \ldots, N$. Given $s_A$, $s_B$ and $\mathbf{Z}^{}(0)=\mathbf{z}$, the random variables
$Z_{ij}^{}(1)$ and $Z_{kl}^{}(1)$ are independent as long as $i\not=k$. In addition, the random vector
$(DZ_{i0}^{}(\mathbf{z}),DZ_{i1}^{}(\mathbf{z}),\ldots,DZ_{iN}^{}(\mathbf{z}))$ has a multinomial conditional distribution with parameters $Dz_i$ and $\left(P_{i0}(\mathbf{z}),P_{i1}(\mathbf{z}),\ldots,P_{iN}(\mathbf{z})\right)$. Therefore, the fractions of demes of the different types at the beginning of generation $1$, given by
\begin{equation}
Z_j(1)= \sum_{i=0}^N Z_{ij}^{}(1)
\end{equation}
for $j=0, 1, \ldots, N$,
satisfy
\begin{subequations}\label{sec1-eq6}
\begin{align}
&E_{\mathbf{z}}\left[Z_j(1)|s_A,s_B\right]= \sum_{i=0}^N E_{\mathbf{z}}\left[ Z_{ij}^{}(1)|s_A,s_B\right]=\sum_{i=0}^{N}z_iP_{ij}(\mathbf{z}),\\
&Var_{\mathbf{z}}\left(Z_j(1)|s_A,s_B\right)=\sum_{i=0}^{N}Var_{\mathbf{z}}\left(Z_{ij}(1)|s_A,s_B\right)
=\frac{1}{D}\sum_{i=0}^{N}z_iP_{ij}(\mathbf{z})\left(1-P_{ij}(\mathbf{z})\right),\\
&Cov_{\mathbf{z}}\left(Z_{j_1}(1),Z_{j_2}(1)|s_A,s_B\right)
=\sum_{i=0}^{N}Cov_{\mathbf{z}}\left(Z_{ij_1}(1),Z_{ij_2}(1)|s_A,s_B\right)
=-\frac{1}{D}\sum_{i=0}^{N}z_iP_{ij_1}(\mathbf{z})P_{ij_2}(\mathbf{z}),
\end{align}
\end{subequations}
for $j, j_1, j_2=0, 1, \ldots, N$ with $j_1\ne j_2$. Here, we denote by $E_{\mathbf{z}}$, $Var_{\mathbf{z}}$ and $Cov_{\mathbf{z}}$ the conditional expectation, variance and covariance, respectively, given that $\mathbf{Z}^{}(0)=\mathbf{z}$.


\section{Approximation by a diffusion process}
In this section, we will deduce a diffusion approximation that relies on the existence of two timescales when the number of demes is large, a long one for the changes in the strategy frequencies given by  $\{X(t)\}_{t\geq 0}$ and a short one for the changes in the deviations of the deme type frequencies from their equilibrium values in an infinite neutral population. These changes are given by the random process
$\left\{\mathbf{Y}(t)=\left(Y_0(t),Y_1(t),\ldots,Y_N(t)\right)\right\}_{t\geq 0}$
where 
\begin{align}
Y_j(t)=Z_j(t)-v_j(X(t)),
\end{align}
for $j=0, 1, \ldots, N$.
Here, the vector $\mathbf{v}(x)=(v_0(x), v_1(x), \ldots,v_N(x))$ represents the equilibrium state of the process $\left\{\mathbf{Z}(t)=\left(Z_0(t),Z_1(t),\ldots,Z_N(t)\right)\right\}_{t\geq 0}$ in the absence of selection when the population is subdivided into an infinite number of demes of the same finite size $N$ and the total frequency of type $A$ is given by the constant $x$. 

More precisely, $\mathbf{v}(x)$ is the solution of the linear system of equations
\begin{equation}\label{sec2-eq1}
v_j(x)=\sum_{i=0}^{N}v_i(x)P_{ij}^{*}(x),
\end{equation}
for $j=0,1,\ldots,N$, where
\begin{equation}\label{sec2-eq2}
P^{*}_{ij}(x)=\binom{N}{j}\left((1-m)\frac{i}{N}+mx\right)^j\left(1-mx-(1-m)\frac{i}{N}\right)^{N-j},
\end{equation}
for $i, j =0, 1, \ldots,N$.
Note that $P^{*}_{ij}(x)$ is the probability of transition from deme type $i$ to deme type $j$ in the absence of selection in the limit of a large number of demes which satisfies
\begin{equation}\label{sec2-eq3}
E_{\mathbf{z}}\Big[P^{}_{ij}(\mathbf{z})\Big]=P^{*}_{ij}(x)+o(1),
\end{equation}
for $i, j =0, 1, \ldots,N$ (see Eq. (\ref{aA-eq3}) in 
Appendix $A$).

A consequence of Eq. (\ref{sec1-eq4}) is that $(P_{ij}(\mathbf{z}))_{j=0}^{N}$ is the probability distribution of a binomial random variable with parameters $N$ and $\tilde{\tilde{x}}_i$. Therefore, we have the identities 
\begin{subequations}\label{sec2-eq4}
\begin{align}
&\sum_{j=0}^{N}jP_{ij}(\mathbf{z})=N\tilde{\tilde{x}}_i,\\
&\sum_{j=0}^{N}j^2P_{ij}(\mathbf{z})=N\tilde{\tilde{x}}_i\left(1-\tilde{\tilde{x}}_i\right)+N^2\tilde{\tilde{x}}_i^2,
\end{align}
\end{subequations}
for $i=0,1,\ldots,N$.
Using these identities, Eq. (\ref{sec1-eq6})  and the fact that
\begin{equation}\label{sec2-eq5}
\sum_{i=0}^{N}z_i\tilde{x}_i=(1-m)\sum_{i=0}^{N}z_i\frac{i}{N}+mx\sum_{i=0}^{N}z_i=(1-m)x+mx=x,
\end{equation}
 we get
\begin{align}\label{sec2-eq6}
E_{\mathbf{z}}\left[X(1)-X(0)\Big|s_A,s_B\right]&=\sum_{j=0}^{N}\frac{j}{N} E_{\mathbf{z}}\left[Z_{j}(1)|s_A,s_B\right]-x\nonumber\\
&=\sum_{j=0}^{N}\frac{j}{N}\sum_{i=0}^{N}z_iP_{ij}(\mathbf{z})-x\nonumber\\
&=\sum_{i=0}^{N}z_i\tilde{\tilde{x}}_i-x.
\end{align}
On the other hand, we have
\begin{equation}\label{sec2-eq7}
E_{\mathbf{z}}\Big[\tilde{\tilde{x}}_i\Big]
=\tilde{x}_i+\frac{(1-\tilde{x}_i)\tilde{x}_i}{ND}\Big[
\mu_A-\mu_B
+(1-\tilde{x}_i)(\sigma_B^2-\sigma_{AB})+\tilde{x}_i(\sigma_{AB}-\sigma_A^2)
\Big]+o(D^{-1}),
\end{equation}
for $i=0,1,\ldots,N$ (see Eq. (\ref{aA-eq2}) in 
Appendix $A$). Combining Eqs. (\ref{sec2-eq6}) and (\ref{sec2-eq7}), 
 the conditional expected change in the frequency of type $A$ from one generation to the next can be written as
\begin{equation}\label{sec2-eq8}
E_{\mathbf{z}}\left[X(1)-X(0)\right]=\frac{1}{ND}M(x,\mathbf{y})+o\left(D^{-1}\right),
\end{equation}
where 
\begin{equation}\label{sec2-eq9}
M(x,\mathbf{y})=\sum_{i=0}^{N}(y_i+ v_i(x))\tilde{x}_i(1-\tilde{x}_i)\Big[\mu_{A}-\mu_{B}+\tilde{x}_i(\sigma_{AB}-\sigma^2_{A})+(1-\tilde{x}_i)\left(\sigma^2_{B}-\sigma_{AB}\right)\Big],
\end{equation}
for $\mathbf{y}=(y_0, y_1, \ldots, y_N)$ with $y_i=z_i-v_i(x)$ for $i=0, 1, \ldots, N$.

Similarly, using Eqs. (\ref{sec1-eq6}) and (\ref{sec2-eq4}), we obtain
\begin{align}\label{sec2-eq8'}
Var_{\mathbf{z}}\left[X(1)-X(0)\Big|s_A,s_B\right]
&=Var_{\mathbf{z}}\left[\sum_{j=0}^{N}\frac{j}{N}Z_{j}(1)\Big|s_A,s_B\right]\nonumber\\
&=\sum_{j=0}^{N}\frac{j^2}{N^2}Var_{\mathbf{z}}\left[Z_{j}(1)\Big|s_A,s_B\right]+\sum_{j_1\not= j_2}\frac{j_1j_2}{N^2}Cov_{\mathbf{z}}\left(Z_{j_1}(1),Z_{j_2}(1)\Big|s_A,s_B\right)\nonumber\\
&=\sum_{j=0}^{N}\frac{j^2}{DN^2}\sum_{i=0}^{N}z_iP_{ij}(\mathbf{z})\left(1-P_{ij}(\mathbf{z})\right)-\sum_{ j_1\not= j_2}\frac{j_1j_2}{DN^2}\sum_{i=0}^{N}z_iP_{ij_1}(\mathbf{z})P_{ij_2}(\mathbf{z})\nonumber\\
&=\frac{1}{D}\sum_{i=0}^{N}z_i\left[\sum_{j=0}^{N}\frac{j^2}{N^2}P_{ij}(\mathbf{z})-\left(\sum_{j=0}^{N}\frac{j}{N}P_{ij}(\mathbf{z})\right)^2\right]\nonumber\\
&=\frac{1}{D}\left[\sum_{i=0}^{N}z_i\left(\frac{\tilde{\tilde{x}}_i(1-\tilde{\tilde{x}}_i)}{N}+\tilde{\tilde{x}}_i^2\right)-\sum_{i=0}^{N}z_i\tilde{\tilde{x}}_i^2\right]\nonumber\\
&=\frac{1}{ND}\sum_{i=0}^{N}z_i\tilde{\tilde{x}}_i(1-\tilde{\tilde{x}}_i).
\end{align}
Therefore, the conditional variance of the change in the frequency of type $A$ from one generation to the next can be expressed as
\begin{align}\label{sec2-eq8''}
&Var_{\mathbf{z}}\left[\left(X(1)-X(0)\right)^2\right]\nonumber\\
&=Var_{\mathbf{z}}\left(
E_{\mathbf{z}}\left[X(1)-X(0)\Big|s_A,s_B\right]
\right)
+
E_{\mathbf{z}}\left(
Var_{\mathbf{z}}\left[X(1)-X(0)\Big|s_A,s_B\right]
\right)\nonumber\\
&=Var_{\mathbf{z}}\left(
\sum_{i=0}^{N}z_i\tilde{\tilde{x}}_i-x
\right)
+
\frac{1}{ND}\sum_{i=0}^{N}z_iE_{\mathbf{z}}\left[\tilde{\tilde{x}}_i(1-\tilde{\tilde{x}}_i)\right]
\nonumber\\
&=
\sum_{i,j=0}^{N}z_iz_jCov_{\mathbf{z}}\left(\tilde{\tilde{x}}_i,\tilde{\tilde{x}}_j\right)+
\frac{1}{ND}\sum_{i=0}^{N}z_iE_{\mathbf{z}}\left[\tilde{\tilde{x}}_i(1-\tilde{\tilde{x}}_i)\right]\nonumber\\
&=\frac{1}{ND}V(x,\mathbf{y})+o\left(D^{-1}\right),
\end{align}
where
\begin{equation}\label{sec2-eq10}
V(x,\mathbf{y})=\left(\sum_{i=0}^{N}(y_i + v_i(x))\tilde{x}_i(1-\tilde{x}_i)\right)^2\left(\sigma_{A}^2+\sigma_B^2-2\sigma_{AB}\right)+\sum_{i=0}^{N}(y_i + v_i(x))\tilde{x}_i(1-\tilde{x}_i)
\end{equation}
using Eqs. (\ref{aA-eq2}), (\ref{aA-eq4}) and (\ref{aA-eq5}) in Appendix $A$.
The functions $M(x,\mathbf{y})$ and $V(x,\mathbf{y})$ represent the population-scaled conditional mean and variance of the change in the frequency of $A$ from one generation to the next in the limit of a large number of demes.

In order to study higher conditional moments of the change in the frequency of $A$, we arbitrarily index the $D_i=Dz_i$ demes of type $i$ at the beginning of generation $0$ with the integers from $1$ to $D_i$ and represent the frequency of $A$ in the deme $(i, d_i)$ at the beginning of generation $1$ by $X_{i,d_i}(1)$ for $d_i=1, \ldots, D_i$ for $i=0, 1, \ldots, N$. Given $s_A$ and $s_B$, these random variables are all independent and satisfy
\begin{equation}\label{sec2-eq12}
E_{\mathbf{z}}\left[X_{i,d_i}(1)\Big|s_A,s_B\right]=\tilde{\tilde{x}}_{i}.
\end{equation}
Moreover, we have
\begin{equation}\label{sec2-eq11}
X(1)=\frac{1}{D}\sum_{i=0}^{N}\sum_{d_i=1}^{D_i}X_{i,d_i}(1).
\end{equation}
Introducing
\begin{equation}
U_{i,d_i}=X_{i,d_i}(1)-\tilde{\tilde{x}}_{i},
\end{equation}
which verifies $\left|U_{i,d_i}\right|\leq1$ and 
$E_{\mathbf{z}}\left[U_{i,d_i}\Big|s_A,s_B\right]=0$, for $d_i=1, \ldots, D_i$ for $i=0, 1, \ldots, N$, we have
\begin{align}\label{}
&E_{\mathbf{z}}\left[\left(X(1)-E_{\mathbf{z}}\left[X(1)\right]\right)^3\Big|s_A,s_B\right]\nonumber\\
&=\frac{1}{D^3}\Bigg[\sum_{(i,d_i)}E_{\mathbf{z}}\left[U_{i,d_i}^3\Big|s_A,s_B\right]
+
3\sum_{(i,d_i)\neq (j,d_j)}E_{\mathbf{z}}\left[U_{i,d_i}^2\Big|s_A,s_B\right]E_{\mathbf{z}}\left[U_{j,d_{j}}\Big|s_A,s_B\right]\nonumber\\
&\quad +\sum_{\substack{(i,d_i), (j,d_j), (k,d_k)\\ \text{all different}}}E_{\mathbf{z}}\left[U_{i,d_i}\Big|s_A,s_B\right]E_{\mathbf{z}}\left[U_{j,d_{j}}\Big|s_A,s_B\right]E_{\mathbf{z}}\left[U_{k,d_{k}}\Big|s_A,s_B\right]
\Bigg]\nonumber\\
&=\frac{1}{D^3}\sum_{(i,d_i)}E_{\mathbf{z}}\left[U_{i,d_i}^3\Big|s_A,s_B\right],
\end{align}
from which we obtain
\begin{align}\label{sec2-eq13}
\left|E_{\mathbf{z}}\left[\left(X(1)-E_{\mathbf{z}}\left[X(1)\right]\right)^3\right]\right|=\Bigg|\frac{1}{D^3}\sum_{(i,d_i)}E_{\mathbf{z}}\left[U_{i,d_i}^3\right]\Bigg|&\leq \frac{D}{D^3}=o(D^{-1}).
\end{align}
Here, we have used the fact that, given $s_A$ and $s_B$, $U_{i,d_i}$ and $U_{j,d_j}$ are conditionally independent as long as $(i,d_i)\neq (j,d_j)$.
Similarly, we have
\begin{align}\label{sec2-eq13'}
&E_{\mathbf{z}}\left[\Big(X(1)-E_{\mathbf{z}}\left[X(1)\right]\Big)^4\Big|s_A,s_B\right]
\nonumber\\
&=\frac{1}{D^4}\Bigg[\sum_{(i,d_i)}E_{\mathbf{z}}\left[U_{i,d_i}^4\Big|s_A,s_B\right]
+
4\sum_{(i,d_i)\neq (j,d_j)}E_{\mathbf{z}}\left[U_{i,d_i}^3\Big|s_A,s_B\right]E_{\mathbf{z}}\left[U_{j,d_{j}}\Big|s_A,s_B\right]\nonumber\\
&\quad
+6\sum_{(i,d_i)\neq (j,d_j)}E_{\mathbf{z}}\left[U_{i,d_i}^2\Big|s_A,s_B\right]E_{\mathbf{z}}\left[U_{j,d_{j}}^2\Big|s_A,s_B\right]\nonumber\\
&\quad +12\sum_{\substack{(i,d_i), (j,d_j), (k,d_k)\\ \text{all different}}}E_{\mathbf{z}}\left[U_{i,d_i}^2\Big|s_A,s_B\right]E_{\mathbf{z}}\left[U_{j,d_{j}}\Big|s_A,s_B\right]E_{\mathbf{z}}\left[U_{k,d_{k}}\Big|s_A,s_B
\right]\nonumber\\
&\quad +\sum_{\substack{(i,d_i), (j,d_j), (k,d_k), (l,d_l)\\ \text{all different}}}E_{\mathbf{z}}\left[U_{i,d_i}\Big|s_A,s_B\right]E_{\mathbf{z}}\left[U_{j,d_{j}}\Big|s_A,s_B\right]E_{\mathbf{z}}\left[U_{k,d_{k}}\Big|s_A,s_B\right]E_{\mathbf{z}}\left[U_{l,d_{l}}\Big|s_A,s_B\right]
\Bigg]\nonumber\\
&=\frac{1}{D^4}\Bigg[\sum_{(i,d_i)}E_{\mathbf{z}}\left[U_{i,d_i}^4\Big|s_A,s_B\right]+
6\sum_{(i,d_i)\neq (j,d_j)}E_{\mathbf{z}}\left[U_{i,d_i}^2\Big|s_A,s_B\right]E_{\mathbf{z}}\left[U_{j,d_{j}}^2\Big|s_A,s_B\right]\Bigg],
\end{align}
from which
\begin{equation}\label{sec2-eq14}
E_{\mathbf{z}}\left[\Big(X(1)-E_{\mathbf{z}}\left[X(1)\right]\Big)^4\right]\leq\frac{1}{D^4}\Big[D+
6D(D-1)\Bigg]=o(D^{-1}).
\end{equation}
Finally, using Eqs. (\ref{sec2-eq7}), (\ref{sec2-eq13}), (\ref{sec2-eq14}) and the fact that $ X(1)-E_{\mathbf{z}}[X(1)]$ is bounded by $1$ in absolute value, we conclude that
\begin{align}\label{sec2-eq16}
E_{\mathbf{z}}\left[\left(X(1)-X(0)\right)^4\right]&=E_{\mathbf{z}}\left[
\left(X(1)-E_{\mathbf{z} }[X(1)]+E_{\mathbf{z}}[X(1)]-X(0)\right)^4
\right]\nonumber\\
&=E_{\mathbf{z}}\left[
\left(X(1)-E_{\mathbf{z}}[X(1)]\right)^4
\right]\nonumber\\
&\quad +
4E_{\mathbf{z}}\left[X(1))-X(0)\right]E_{\mathbf{z}}\left[
\left(X(1)-E_{\mathbf{z}}[X(1)]\right)^3
\right]\nonumber\\
&\quad+o\left(\left|E_{\mathbf{z}}\left[X(1))-X(0)\right]\right|\right)
\end{align}
is a function $o(D^{-1})$.

Now, we will study the change 
\begin{align}\label{sec2-eq17.0}
Y_j(1)-Y_j(0)= Z_j(1)-Z_j(0) - v_j(X(1))+v_j(X(0)),
\end{align}
where $v_j$ is defined in Eq. (\ref{sec2-eq1}), 
for $j=0, 1, \ldots, N$. We will make use of the important property conjectured in Wakeley (2003) and shown in Lessard (2007) that
\begin{equation}\label{sec2-eq17}
v_j(X(1))=v_j(X(0))+\sum_{k=1}^{N}r_k\left(X(1)-X(0)\right)^k,
\end{equation} 
where $r_k$ for $k=1, \ldots, N$ are constants that depend only on $N$, $m$ and $x$ (see Eq. (7.7) in Lessard, 2009, for further details). In addition, since $|X(1)-X(0)|\leq1$, we have
\begin{equation}
E_{\mathbf{z}}\left[\left|X(1)-X(0)\right|^k\right]
\leq E_{\mathbf{z}}\left[\left|X(1)-X(0)\right|^2\right],
\end{equation}
from which
 \begin{equation}\label{sec2-eq17.1}
E_{\mathbf{z}}\left[\left(X(1)-X(0)\right)^k\right]= o(1)
\end{equation}
owing to Eqs. (\ref{sec2-eq7}) and (\ref{sec2-eq9}), 
for $k\geq 2$. 
Using this  result and Eq. (\ref{sec2-eq7}) again, we obtain
\begin{equation}\label{sec2-eq18}
E_{\mathbf{z}}\left[v_j(X(1))\right]=v_j(x)+\sum_{k=1}^{N}r_k E_{\mathbf{z}}\left[\left(X(1)-X(0)\right)^k\right]=v_j(x)+o(1),
\end{equation} 
for $j=0, 1, \ldots, N$.
 On the other hand, using Eqs. (\ref{sec1-eq6}) and (\ref{sec2-eq1}) with $z_i=y_i+v_i(x)$ for $i=0, 1, \ldots, N$ yields
\begin{align}\label{sec2-eq19}
E_{\mathbf{z}}\left[Z_j(1)\right]&=\sum_{i=0}^{N}z_iE_{\mathbf{z}}\left[P_{ij}(\mathbf{z})\right]=\sum_{i=0}^{N}(y_i+v_i(x))P_{ij}^*(x)+o(1)=\sum_{i=0}^{N}y_iP_{ij}^*(x)+v_j(x)+o(1).
\end{align}
Therefore, combining Eqs. (\ref{sec2-eq18}) and (\ref{sec2-eq19}), we get
\begin{equation}\label{sec2-eq20}
\begin{split}
E_{\mathbf{z}}\left[Y_j(1)-Y_j(0)\right]&=E_{\mathbf{z}}\left[Z_j(1)\right]-E_{\mathbf{z}}\left[v_j(X(1))\right]-y_j=c_j(x,\mathbf{y})+o\left(1\right),
\end{split}
\end{equation}
where
\begin{equation}\label{sec2-eq21}
c_j(x,\mathbf{y})=\sum_{i=0}^{N}y_iP_{ij}^*(x)-y_j
\end{equation}
with $y_j=z_j-v_j(x)$, for $j=0, 1, \ldots, N$.

We will establish next that 
\begin{equation}\label{sec2-eq211}
Var_{\mathbf{z}}\left[Y_j(1)-Y_j(0)\right]=o\left(1\right),
\end{equation}
for $j=0, 1, \ldots, N$.
Note first that, for any two random variables $U_1$ and $U_2$, we have
\begin{equation}\label{sec2-eq23}
Var(U_1+U_2)=Var(U_1)+Var(U_2)+2Cov(U_1,U_2)\leq 2Var(U_1)+2Var(U_2),
\end{equation}
owing to the inequalities
\begin{equation}\label{sec2-eq22}
Cov(U_1,U_2)\leq \sqrt{Var(U_1)}\sqrt{Var(U_2)}\leq \frac{Var(U_1)+Var(U_2)}{2}.
\end{equation} 
Therefore, from Eq. (\ref{sec2-eq17.0}), we have
\begin{equation}\label{sec2-eq24}
Var_{\mathbf{z}}\left[ Y_j(1)-Y_j(0)\right]=Var_{\mathbf{z}}\left[Z_j(1)-v_j(X(1))\right]\leq 
2Var_{\mathbf{z}}\left[Z_j(1)\right]+2Var_{\mathbf{z}}\left[v_j(X(1))\right].
\end{equation}
In order to establish Eq. (\ref{sec2-eq211}) as $D$ goes to infinity, it is enough to have $Var_{\mathbf{z}}\left[Z_j(1)\right]=o(1)$, which is a direct consequence of Eq. (\ref{sec1-eq6}), and 
$Var_{\mathbf{z}}\left[v_j(X(1))\right]=o(1)$. As a matter of fact, this variance is given by
\begin{align}\label{sec2-eq25}
Var_{\mathbf{z}}\left[v_j(X(1))\right]
&=Var_{\mathbf{z}}\left[\sum_{k=1}^{N}r_k\left(X(1)-X(0)\right)^k\right]\nonumber\\
&=\sum_{k_1,k_2=1}^{N}r_{k_1}r_{k_2}Cov_{\mathbf{z}}\left[\left(X(1)-X(0)\right)^{k_1},\left(X(1)-X(0)\right)^{k_2}\right]\nonumber\\
&=\sum_{k_1,k_2=1}^{N}r_{k_1}r_{k_2}\Bigg(
E_{\mathbf{z}}\left[\left(X(1)-X(0)\right)^{k_1+k_2}\right]\nonumber\\
&\quad\quad-E_{\mathbf{z}}\left[\left(X(1)-X(0)\right)^{k_1}\right]E_{\mathbf{z}}\left[\left(X(1)-X(0)\right)^{k_2}\right]
\Bigg),
\end{align}
which is a function $o(1)$ owing to Eqs. (\ref{sec2-eq7}) and (\ref{sec2-eq17.1}). This completes the proof of Eq. (\ref{sec2-eq211}).

Finally, we consider the deterministic difference equation
\begin{equation}\label{sec2-eq27}
\mathbf{Y}(k+1,x,\mathbf{y})-\mathbf{Y}(k,x,\mathbf{y})=\mathbf{c}(x,\mathbf{Y}(k,x,\mathbf{y})),
\end{equation}
for every integer $k \geq 0$ with initial condition $\mathbf{Y}(0,x,\mathbf{y})=\mathbf{y}$, where 
\begin{align}
\mathbf{c}(x,\mathbf{y})=(c_0(x,\mathbf{y}), c_1(x,\mathbf{y}), \ldots, c_N(x,\mathbf{y}))
\end{align}
is defined by Eq. (\ref{sec2-eq21}). Owing to Eq. (\ref{sec2-eq1}), this equation 
can be rewritten as 
\begin{equation}\label{sec2-eq28}
\mathbf{Z}(k+1,x,\mathbf{y})=\mathbf{Z}(k,x,\mathbf{y})P^{*}(x),
\end{equation}
where $\mathbf{Z}(k,x,\mathbf{y})=\mathbf{Y}(k,x,\mathbf{y})+\mathbf{v}(x)$ for $k\geq 0$ with $\mathbf{v}(x)=(v_0(x), v_1(x), \ldots, v_N(x))$, and 
$P^{*}(x)=(P_{ij}^{*}(x))_{i, j=0}^N$ is a transition matrix of an irreducible Markov chain on a finite state space with $\mathbf{v}(x)$ as stationary distribution. The ergodic theorem ensures that
\begin{equation}\label{sec2-eq29}
\lim_{k\rightarrow\infty}\mathbf{Z}(k,x,\mathbf{y})=\mathbf{v}(x),
\end{equation}
which is equivalent to 
\begin{equation}
\lim_{k\rightarrow\infty}\mathbf{Y}(k,x,\mathbf{y})=\mathbf{0}=\mathbf{c}(x,\mathbf{0}).
\end{equation}

Let us summarize our findings:
\begin{subequations}
\begin{align}
E_{\mathbf{z}}\left[X(1)-X(0)\right]&=\frac{1}{ND}M(x,\mathbf{y})+o\left(D^{-1}\right),\\
E_{\mathbf{z}}\left[\left(X(1)-X(0)\right)^2\right]&=\frac{1}{ND}V(x,\mathbf{y})+o\left(D^{-1}\right),\\
E_{\mathbf{z}}\left[\left( X(1)-X(0)\right)^4\right]&=o(D^{-1}),\\
E_{\mathbf{z}}\left[Y_j(1)-Y_j(0)\right]&=c_j(x,\mathbf{y})+o\left(1\right),\\
Var_{\mathbf{z}}\left[Y_j(1)-Y_j(0)\right]&=o\left(1\right),
\end{align}
\end{subequations}
for $j=0, 1, \ldots, N$, where $M(x,\mathbf{y})$, $V(x,\mathbf{y})$ and $\mathbf{c}(x,\mathbf{y})=(c_0(x,\mathbf{y}), c_1(x,\mathbf{y}), \ldots, c_N(x,\mathbf{y}))$ are smooth functions as defined in Eqs. (\ref{sec2-eq1}), (\ref{sec2-eq8}), (\ref{sec2-eq10}) and (\ref{sec2-eq21}). Moreover, $\mathbf{c}(x,\mathbf{y})=\mathbf{0} $ is a solution of the recurrence equation
\begin{equation}\label{sec2-eq27}
\mathbf{Y}(k+1,x,\mathbf{y})-\mathbf{Y}(k,x,\mathbf{y})=\mathbf{c}(x,\mathbf{Y}(k,x,\mathbf{y}))
\end{equation}
that 
is asymptotically stable given any initial state $\mathbf{Y}(0,x,\mathbf{y})=\mathbf{y}$. Actually, all these conditions are uniform with respect to the population state and the additional condition  $\mu(0,\mathbf{y})=0$ ensures the existence of a strongly continuous semigroup corresponding to a diffusion process (Ethier, 1976). Owing to  Theorem 3.3 in Ethier and Nagylaki (1980), we can conclude as below.
\begin{Proposition}
Let $X(\lfloor ND\tau\rfloor)$ be the frequency of type $A$ at time $\tau\geq0$ in number of $ND$ generations in a haploid population subdivided into $D$ demes, where each deme has $N$ individuals with viability coefficients satisfying (\ref{sec1-eq1}) and a fraction $m$ of offspring disperse uniformly among the demes. Here,
$\lfloor x\rfloor$ designates the integer part of $x$. Then, as $D$ goes to infinity, the process $\{X(\lfloor ND\tau\rfloor)\}_{\tau\geq0}$ converges
in distribution to a diffusion process $\{X^\ast(\tau)\}_{\tau\geq0}$ on $[0, 1]$ whose generator is 
\begin{equation}\label{sec3-eq1}
\mathcal{L}=\frac{1}{2}V(x,\mathbf{0})\frac{d^2}{dx^2}+M(x,\mathbf{0})\frac{d}{dx}.
\end{equation}
Here, the infinitesimal mean and variance are given by
\begin{subequations}\label{sec3-eq2}
\begin{align}
&M(x,\mathbf{0})=(1-f)x(1-x)\Big[\mu_A-\mu_B+C_1(\sigma^2_B-\sigma_{AB})-C_2(\sigma^2_A-\sigma_{AB})\Big],\\
&V(x,\mathbf{0})=(1-f)x(1-x)\Big[1+(1-f)x(1-x)\left(\sigma_{A}^2+\sigma_B^2-2\sigma_{AB}\right)\Big],
\end{align}
\end{subequations}
where
\begin{subequations}
\begin{align}
&f=\frac{(1-m)^2}{(1-m)^2+mN(2-m)},\label{fixationindex}\\
&C_1=\frac{(1-m)^2N[3-m(3-m)(1+x)]+(2-m)\left[mN^2x(3-3m+m^2)-(1-m)^2\right]}{[N^2-(1-m)^3(N-1)(N-2)](2-m)},\\
&C_2=\frac{(1-m)^2N[3-m(3-m)(2-x)]+(2-m)\left[mN^2(1-x)(3-3m+m^2)-(1-m)^2\right]}{[N^2-(1-m)^3(N-1)(N-2)](2-m)}.
\end{align}
\end{subequations}
This is true for any fixed  deme size $N\geq 2$ and any dispersal fraction $m\in(0,1)$.
\end{Proposition}
The derivation of 
$M(x,\mathbf{0})$ and $V(x,\mathbf{0})$ is relegated  to Appendix $B$. Note that $f$ given above is well known in the literature as the fixation index for a haploid population subdivided into an infinite number of groups of  fixed size $N$ with a constant dispersal fraction $m$ of offspring (see, e.g., Wakeley, 2003).
\section{Fixation probability}
In the absence of mutation, the frequency of type $A$ in the population is a Markov chain with two fixation states, $x=1$ when all individuals are of type $A$, and  $x=0$ when they are all of type $B$. 
 Let 
$P_1(p,\tau)$ be the probability for the diffusion process $\{X^\ast(\tau)\}_{\tau\geq0}$ to reach $x=1$ at or before time $\tau$ starting from $x=p$ at time $0$, so that
$X(\tau)=1$ given that $X(0)=p$. Note that this probability satisfies the backward Kolmogov equation, that is,
\begin{equation}\label{sec4-eq1}
-\frac{\partial P_1(p,\tau)}{\partial \tau}=M(x,\mathbf{0})\frac{\partial P_1(p,\tau)}{\partial p}+
\frac{V(x,\mathbf{0})}{2}\frac{\partial^2P_1(p,\tau)}{\partial p^2},
\end{equation}
with the boundary conditions $P_1(0,\tau)=0$ and $P_1(1,\tau)=1$. See Karlin and Taylor \cite{KT1981} or Ewens \cite{E2004} for more details. 

Let $F_A(p)=\lim_{\tau \rightarrow\infty}P_1(p,\tau)$ be the fixation probability of type $A$ given that its initial frequency is $p$. Then, equation (\ref{sec4-eq1}) yields
\begin{equation}\label{sec4-eq2}
0=M(x,\mathbf{0})\frac{d F_A(p)}{d p}+
\frac{V(x,\mathbf{0})}{2}\frac{d^2F_A(p)}{d p^2},
\end{equation}
with the boundary conditions $F_A(0)=0$ and $F_A(1)=1$. This equation has the exact solution
\begin{equation}\label{sec4-eq3}
F_A(p)=\frac{\int_{0}^pS(y)dy}{\int_{0}^1S(y)dy},
\end{equation}
where 
\begin{equation}\label{sec4-eq4}
S(y)=\exp\left\{-2\int_{0}^{y}\frac{M(x,\mathbf{0})}{V(x,\mathbf{0})}dx\right\}.
\end{equation}
See the same references as above.

Now, define $F_A$ as the fixation probability of $A$ introduced as a single mutant in a population of $B$ individuals distributed in $D$ demes of size $N$ as in the previous sections.  For $D$ large enough, the discrete-time process $\{X(t)\}_{t\geq0}$ for the frequency of $A$ can be approximated by the diffusion process $\{X^\ast(\tau)\}_{\tau\geq0}$ given in Proposition 1, from which we have
\begin{equation}\label{sec4-eq5}
F_A\approx F_A\left(\frac{1}{ND}\right)=\frac{\int_{0}^{(ND)^{-1}}S(y)dy}{\int_{0}^1S(y)dy}\approx\frac{1}{ND\int_{0}^1S(y)dy},
\end{equation}
since
\begin{equation}\label{sec4-eq6}
\int_{0}^{(ND)^{-1}}S(y)dy\approx\frac{S(0)}{ND}=\frac{1}{ND},
\end{equation}
where $S(y)$ is given in Eq. (\ref{sec4-eq4}).

\section{Large deme size and small dispersal fraction}
In this section, we suppose that the deme size $N$ is large and the dispersal fraction $m$ is small. More precisely, we  let $N\rightarrow\infty$ and $Nm\rightarrow\nu$  in the model of section 2. Here, the parameter $\nu$ corresponds to a \emph{deme-scaled dispersal rate}. 
Note that we have the approximations
\begin{subequations}\label{sec5-eq1}
\begin{align}
&(1-m)^2+mN(2-m)=1+2\nu+o(1),\\
&N^2- (1-m)^3(N-1)(N-2)=3N(1+\nu)+o(N),\\
&(1-m)^2N[3-m(3-m)(1+x)]+(2-m)\left[mN^2x(3-3m+m^2)-(1-m)^2\right]\nonumber\\
&\quad\quad\quad\quad\quad\quad\quad\quad\quad\quad\quad\quad\quad\,=3N(1+2\nu x)+o(N),\\
&(1-m)^2N[3-m(3-m)(2-x)]+(2-m)\left[mN^2(1-x)(3-3m+m^2)-(1-m)^2\right]\nonumber\\
&\quad\quad\quad\quad\quad\quad\quad\quad\quad\quad\quad\quad\quad\,=3N(1+2\nu(1-x))+o(N),
\end{align}
\end{subequations}
from which we obtain 
\begin{subequations}\label{sec5-eq2}
\begin{align}
f&=\frac{1}{1+2\nu}+o(1),\\
C_1&=\frac{1+2\nu x}{2(1+\nu)}+o(1),\\
C_2&=\frac{1+2\nu(1-x)}{2(1+\nu)}+o(1).
\end{align}
\end{subequations}
Then, in the limit of a large deme size, the infinitesimal mean and variance in Proposition 1 become
\begin{subequations}\label{sec5-eq4}
\begin{align}
M(x,\mathbf{0})&=\frac{2\nu x(1-x)}{1+2\nu}\left[\mu_A-\mu_B+\frac{\sigma_{AB}-\sigma_A^2}{2(1+\nu)}(1+2\nu x)
+\frac{\sigma_B^2-\sigma_{AB}}{2(1+\nu)}(1+2\nu(1-x))\right],\\
V(x,\mathbf{0})&=\frac{2\nu x(1-x)}{1+2\nu}\left[1+\frac{2\nu x(1-x)}{1+2\nu}(\sigma_{A}^2+\sigma_B^2-2\sigma_{AB})\right].
\end{align}
\end{subequations}
In this case, Eq. (\ref{sec4-eq4}) yields
\begin{equation}\label{sec5-eq7}
S(y)=\exp\left\{- 2\int_{0}^{y}g(x)dx\right\},
\end{equation}
where 
\begin{equation}\label{sec5-eq8}
\begin{split}
g(x)&=\frac{\mu_A-\mu_B+\dfrac{\sigma_{AB}-\sigma_A^2}{2(1+\nu)}(1+2\nu x)+\dfrac{\sigma_B^2-\sigma_{AB}}{2(1+\nu)}(1+2\nu(1-x))}
{1+\dfrac{2\nu x(1-x)}{1+2\nu}(\sigma_{A}^2+\sigma_B^2-2\sigma_{AB})}.
\end{split}
\end{equation}

\subsection{Uncorrelated viability coefficients}

In this subsection, we suppose that the viability coefficients $s_A$ and $s_B$ are uncorrelated. In this case, we have
\begin{equation}
\frac{\partial g}{\partial \sigma_B^2}=\dfrac{\dfrac{2\nu x(1-x)}{1+2\nu}(\mu_B-\mu_A)+\dfrac{1+2\nu(1-x)}{2(1+\nu)}(1+\sigma_A^2)}{\left(1+\dfrac{2\nu x(1-x)}{1+2\nu}(\sigma_{A}^2+\sigma_B^2)\right)^2}>0,
\end{equation}
at least as long as $\mu_A<\mu_B$.
In this case, increasing the population-scaled variance of the viability coefficient of type $B$ will increase the fixation probability $F_A$. 
\paragraph{Result 1.}
\emph{If the viability coefficients are uncorrelated and their population-scaled means satisfy $\mu_A<\mu_B$, then increasing the population-scaled variance of the viability coefficient of type $B$ will increase the fixation probability of type $A$ introduced as a single mutant in an all $B$-population.
}

Now, assume that 
\begin{equation}\label{cond1}
\mu_B-\mu_A<\frac{1+2\nu}{2(1+\nu)}(1+\sigma^2_B).
\end{equation}
This leads to 
\begin{equation}
h(x)=(1+2\nu)(1+(1-x)\sigma^2_B)+2(1+\nu)(1-x)(\mu_A-\mu_B)>0,
\end{equation}
which implies that 
\begin{equation}
\frac{\partial g}{\partial \sigma_A^2}=-\dfrac{\dfrac{1}{2(1+\nu)}+\dfrac{2\nu xh(x)}{2(1+2\nu)(1+\nu)}}{\Big(1+\cfrac{2\nu x(1-x)}{1+2\nu}(\sigma_{A}^2+\sigma_B^2)\Big)^2}<0,
\end{equation}
for $x\in[0,1]$. This implies that an increase in the population-scaled variance of the viability coefficient of type $A$ will decrease the fixation probability  $F_A$.
\paragraph{Result 2.}
\emph{If the viability coefficients are uncorrelated and if the inequality in Eq. (\ref{cond1}) is satisfied, then increasing the population-scaled variance of the viability coefficient of type $A$ will decrease the fixation probability of type $A$ introduced as a single mutant in an all $B$-population.
}


\subsection{Case: $\sigma_{A}^2=\sigma_{B}^2=\sigma_{AB}$}
If $\sigma_{A}^2=\sigma_{B}^2=\sigma_{AB}$, which includes the case where $s_A$ and $s_B$ are deterministic, then we have
\begin{equation}\label{sec5-eq11}
S(y)=\exp\left\{-2\int_{0}^{y}\frac{M(x,\mathbf{0})}{V(x,\mathbf{0})}dx\right\}=e^{-2(\mu_A-\mu_B)y}.
\end{equation}
Therefore, the fixation probability of type $A$ introduced as a single mutant in a population of $B$ individuals can be approximated either as
\begin{equation}\label{sec5-eq12}
F_A\approx\frac{1}{ND}
\end{equation}
if the viability coefficients of both types have the same population-scaled mean, that is, $\mu_A=\mu_B$, or as
\begin{equation}\label{sec5-eq13}
F_A\approx\frac{2(\mu_A-\mu_B)}{ND(1-e^{-2(\mu_A-\mu_B)})}
\end{equation}
if their  population-scaled means are different. Note that $\nu$ has no effect on the approximation of $F_A$. By symmetry, $\nu$ has no effect on the approximation of $F_B$.

Moreover, since $x>1+e^{-x}$ if $x>0$ and $x<1+e^{-x}$ if $x<0$, we conclude that selection favours the fixation of type $A$ in the sense that $F_A>(ND)^{-1}$ if $\mu_A>\mu_B$. On the other hand, if $\mu_A<\mu_B$, then selection disfavours the fixation of $A$ in the sense that $F_A<(ND)^{-1}$. 

By symmetry, the fixation probability of type $B$ introduced as a single mutant in a population of $A$ individuals can be approximated as
\begin{equation}\label{sec5-eq14}
F_B\approx\frac{2(\mu_B-\mu_A)}{ND(1-e^{-2(\mu_B-\mu_A)})},
\end{equation}
from which $F_B>(ND)^{-1}$, and consequently selection favours the fixation of $B$, if $\mu_B>\mu_A$,  while selection disfavours the fixation of $B$ if $\mu_B<\mu_A$.

Let us summarize our findings in this subsection. 

\paragraph{Result 3.}
\emph{If $\sigma_{A}^2=\sigma_{B}^2=\sigma_{AB}$, then selection fully favours the fixation of type $A$ in the sense that $F_A>(ND)^{-1}>F_B$ as long as
$\mu_A>\mu_B$. Dispersal has no effect on which strategy is favoured by selection.
}

\subsection{Low deme-scaled dispersal rate}
It is of interest to consider the effect of a high level of relatedness within demes. This occurs in the case of a low deme-scaled dispersal rate, that is, when $\nu\rightarrow 0$. Under this scenario, we obtain
\begin{equation}\label{sec5-eq16}
g(x)=\mu_A-\mu_B+\frac{\sigma_B^2-\sigma_A^2}{2}.
\end{equation}
In this case, the fixation probability of $A$ introduced as a single mutant can be approximated as
\begin{equation}\label{sec5-eq17}
F_A\approx\frac{2(\mu_A-\mu_B)+\sigma_B^2-\sigma_A^2}{ND\Big(1-e^{-(2(\mu_A-\mu_B)+\sigma_B^2-\sigma_A^2)}\Big)}.
\end{equation}
By symmetry, the corresponding fixation probability for $B$ is approximated as
\begin{equation}\label{sec5-eq18}
F_B\approx\frac{2(\mu_B-\mu_A)+\sigma_A^2-\sigma_B^2}{ND(1-e^{-[2(\mu_B-\mu_A)+\sigma_A^2-\sigma_B^2]})}.
\end{equation}
Note that $\sigma_{AB}$ does not have any effect on the approximations of $F_A$ and $F_B$.
Then, the condition for $F_A>(ND)^{-1}$, which is exactly the condition for $F_B<(ND)^{-1}$, is
\begin{equation}\label{sec5-eq19}
\mu_A-\frac{\sigma_A^2}{2}>\mu_B-\frac{\sigma_B^2}{2}.
\end{equation}
\textcolor{black}{The quantities $\mu_A-\sigma_A^2/2$ and $\mu_B-\sigma_B^2/2$ correspond to the population-scaled geometric mean viability coefficient of types $A$ and $B$, respectively (see Gillespie \cite{G1973}).
Condition (\ref{sec5-eq19}) implies that selection will be favouring the fixation of the type for which this geometric mean is the highest, while disfavouring the fixation of the other type.}

\textcolor{black}{Before going further, let us provide an explanation for the above terminology. The geometric mean of a random variable $U$ can be defined as
\begin{equation}
GM_U=e^{E[\log(U)]}.
\end{equation}
Once we apply the Taylor expansion of $\log(1+s_i)$ with respect to the viability coefficient $s_i$ for type $i$, we obtain
\begin{equation}
E[\log(1+s_i)]=E[s_i]-\frac{E[s^2_i]}{2}+o\left(D^{-1}\right)=\frac{1}{ND}\left(\mu_i-\frac{\sigma_i^2}{2}\right)+o\left(D^{-1}\right).
\end{equation}
Consequently, the geometric mean of $1+s_i$ can be approximated as
\begin{equation}
GM_{1+s_i}\approx 1+\frac{1}{ND}\left(\mu_i-\frac{\sigma_i^2}{2}\right).
\end{equation}
Therefore, we can refer to $\mu_i-\sigma_i^2/2$ as the population-scaled geometric mean viability coefficient of type $i$.}

 In the limit of a low deme-scaled dispersal rate, the effect of dispersal on the evolutionary process  is negligible. The evolutionary process in each deme behaves as in a well-mixed population where fixation will occur first in each deme to become only of type $A$ or only of type $B$. Then, within each deme, there is a competition between individuals of the same type. This explains why the population-scaled covariance $\sigma_{AB}$ has no effect on which strategy is favoured by selection with respect to fixation probability. The only competition will hold between demes to overtake the population. 

\subsection{Small population-scaled means, variances and covariance}
Assume that $\mu_A$ and $\mu_B$ as well as $\sigma_A^2$, $\sigma_B^2$ and $\sigma_{AB}$, are of the same small enough order. Then, we have the approximation
\begin{equation}
g(x)\approx \mu_A-\mu_B+\dfrac{\sigma_{AB}-\sigma_A^2}{2(1+\nu)}(1+2\nu x)+\dfrac{\sigma_B^2-\sigma_{AB}}{2(1+\nu)}(1+2\nu(1-x)),
\end{equation}
from which we obtain 
\begin{align}
S(y)&=\exp\left(-2\int_{0}^{y}g(x)dx\right)\nonumber\\
&\approx 1-2\int_{0}^{y}g(x)dx\nonumber\\
&\approx 1- \left[2y\left(\mu_A-\mu_B\right)+\frac{\sigma_{AB}-\sigma_A^2}{1+\nu}(y+\nu y^2)+
\frac{\sigma_B^2-\sigma_{AB}}{1+\nu}\left((1+2\nu)y-\nu y^2\right)\right].
\end{align}
Here, we have used the approximation $e^{u}\approx 1+u$ for $u$ small enough.
Substituting this expression of $S(y)$ in Eq. (\ref{sec4-eq5}), the fixation probability  $F_A$ can be approximated as
\begin{equation}\label{FAweak}
F_A\approx \frac{1}{ND}+\frac{1}{ND}\left[
\mu_A-\mu_B+\frac{3+2\nu}{6(1+\nu)}(\sigma_{AB}-\sigma_A^2)+\frac{3+4\nu}{6(1+\nu)}(\sigma_B^2-\sigma_{AB})
\right].
\end{equation}
We deduce that selection favours the fixation of type $A$ introduced as a single mutant in an all $B$-population as long as
\begin{equation}
\mu_A-\mu_B+\frac{3+4\nu}{6(1+\nu)}\sigma_B^2 -\frac{3+2\nu}{6(1+\nu)}\sigma_A^2-\frac{2\nu}{6(1+\nu)}\sigma_{AB}>0.
\end{equation}
Decreasing the population-scaled variance of the viability coefficient of type $A$ or increasing the population-scaled variance of the viability coefficient of type $B$ will increase the fixation probability $F_A$. By symmetry, this will decrease the fixation probability  $F_B$. Note that an increase in the population-scaled covariance between the viability coefficient of type $A$ and the viability coefficient of type $B$ will decrease both fixation probabilities, $F_A$ and $F_B$.

Another important point is the effect of $\nu$ on $F_A$. Note that 
\begin{equation}
\frac{d}{d\nu}\left(
\frac{3+4\nu}{6(1+\nu)}\sigma_B^2 -\frac{3+2\nu}{6(1+\nu)}\sigma_A^2-\frac{2\nu}{6(1+\nu)}\sigma_{AB}
\right)
=\frac{\sigma_B^2+\sigma_A^2-2\sigma_{AB}}{6(1+\nu)^2}\geq 0.
\end{equation}
Then, increasing the deme-scaled dispersal rate will increase the fixation probability $F_A$, and also, by symmetry, the fixation probability $F_B$.

Note that the coefficients of $\sigma_A^2$ and $\sigma_B^2$ in Eq. (\ref{FAweak}) 
are increasing functions of $\nu$, while the coefficient of $\sigma_{AB}$ is a decreasing function. This means that an increase in the deme-scaled dispersal rate will increase the weights of the population-scaled variances in the approximation of $F_A$, while it will decrease the weight of the population-scaled covariance.

As $\nu\rightarrow 0$, we have the approximation
\begin{equation}
F_A\approx \frac{1}{ND}+\frac{1}{ND}\left[
\mu_A-\mu_B+\frac{\sigma_B^2}{2}-\frac{\sigma_A^2}{2}
\right],
\end{equation}
while as
$\nu\rightarrow \infty$, we have rather
\begin{equation}
F_A\approx \frac{1}{ND}+\frac{1}{ND}\left[
\mu_A-\mu_B+\frac{2}{3}\sigma_B^2-\frac{1}{3}\sigma_A^2-\frac{1}{3}\sigma_{AB}
\right].
\end{equation}
In the former case, selection favours the fixation of type $A$ introduced as a single mutant as long as $\mu_A-\sigma_A^2/2>\mu_B-\sigma_B^2/2$ in agreement with the previous subsection in the case of a low deme-scaled dispersal rate. In the latter case, however, the condition becomes
\begin{equation}
\mu_A-\mu_B+\frac{2}{3}\sigma_B^2-\frac{1}{3}\sigma_A^2-\frac{1}{3}\sigma_{AB}>0.
\end{equation}
The same condition was obtained in the case of a very large well-mixed population that evolves according to a Moran model (see Eq. ($38$) in Kroumi \textit{et al.}, 2021,  for $\eta_1=\eta_2=s_A$ and $\eta_3=\eta_4=s_B$).
Therefore, under a high deme-scaled dispersal rate, the evolutionary process in the island model is analogous to the one in a well-mixed population.

It is interesting to note that $F_A>F_B$ if and only if $\mu_A-\sigma_A^2/2>\mu_B-\sigma_B^2/2$ irrespective of the deme-scaled dispersal rate.
\section{Discussion}

Previous studies examining the island model primarily focused on constant viability selection. In this paper, we have studied a haploid population subdivided into $D$ demes, each of size $N\geq 2$, in which an individual can be either of type $A$, with viability coefficient $s_A$, or of type $B$, with viability coefficient $s_B$. We have assumed that the viability coefficients fluctuate in a random manner from one generation to the next such that the means, variances, and covariance are inversely proportional to $ND$. The reproduction process in each deme is according to a Wright-Fisher sampling procedure with a constant uniform dispersal fraction of offspring, denoted by $m$, before selection.

Similarly to Wakeley \cite{W2003} with the ancillary result in Lessard \cite{L2007}, we have established that the discrete-time Markov chain, given by the frequency of demes for each type and the total frequency of type $A$ in the population, meets the key conditions given in Ethier and Nagylaki \cite{EN1980} as $D$ approaches infinity. This allows us to approximate the total frequency of type $A$ with a diffusion process. Using this approach, it becomes possible to calculate the probability that type $A$ introduced as a single mutant, represented by $F_A$,  becomes fixed in order to examine the influence of stochastic viability coefficients on the evolutionary dynamics.

If the viability coefficients are deterministic, type $A$ is favoured by selection in the sense that $F_A> (ND)^{-1}$ only when the population-scaled means, which correspond here to the population-scaled values, satisfy $\mu_A>\mu_B$, just as it is in a well-mixed population, where the deme-scaled dispersal rate $\nu=\lim_{N\rightarrow\infty}Nm$ has no impact on the fixation probability. 
These conclusions are no more valid with stochastic viability coefficients. 
For instance, if $\mu_A<\mu_B$ and the viability coefficients are uncorrelated, increasing the population-scaled variance of the viability coefficient of type $B$ will increase the fixation probability of type $A$. Furthermore, decreasing the population-scaled variance of the viability coefficient of type $A$ will also increase $F_A$ under additional conditions.  Therefore, a stochastic environment has the potential to reverse the favoured strategy. Similar results have been proven in well-mixed populations with fluctuations in viability coefficients in a population genetic framework (Kimura \cite{K1954}, Gillespie \cite{G1973,G1974}, Karlin and Levikson \cite{KLe1974}, Karlin and Liberman \cite{KLi1974}, Avery \cite{A1977}) or in the payoffs in the context of evolutionary game theory (Li and Lessard \cite{LL2020}, Kroumi and Lessard \cite{KL2021,KL2022}, Kroumi \textit{et al.} \cite{KMLL2021,KML2022}).  It is important to note that if the population-scaled variances and covariance, $\sigma_A^2,$ $\sigma_B^2$, and $\sigma_{AB}$, are of the same magnitude, the evolutionary process behaves as it would under deterministic viability coefficients.

Our results show that by scaling the viability coefficients in terms of $ND/(1-f)$ generations instead of $ND$ generations, where $f$ is the fixation index given in Eq. (\ref{fixationindex}), it is possible to express the infinitesimal mean and variance of the limiting diffusion process using measures of relatedness in an infinite population under neutrality. More precisely, we have
\begin{subequations}
\begin{align}
M(x,\mathbf{0})&=x(1-x)\left[\mu_A-\frac{\sigma_A^2}{2}-\left(\mu_B-\frac{\sigma_B^2}{2}\right)+\left(1-\frac{f_{IJK}}{f_{IJ}}\right)\sigma^2\left(\frac{1}{2}-x\right)\right],\\
V(x,\mathbf{0})&=x(1-x)\left[1+\left(1-f_{IJ}\right)\sigma^2x(1-x)\right],
\end{align}
\end{subequations}
 where $\sigma^2=\sigma_A^2+\sigma_B^2-2\sigma_{AB}$, while $I$, $J$, and $K$ are three offspring randomly selected from the same deme in an infinite neutral population and
\begin{subequations}\label{d-eq1}
\begin{align}
&f_{IJ}=\mathbb{P}\left(I\equiv J\right)=\frac{1}{1+2\nu},\\
&f_{IJK}=\mathbb{P}\left(I\equiv J\equiv k\right)=\frac{1}{(1+\nu)(1+2\nu)}.
\end{align}
\end{subequations}
Here, the symbol $\equiv$ means that the individuals are identical by descent and $\not\equiv$ means that the individuals are not identical by descent. For the computation of $f_{IJ}$ and $f_{IJK}$, see Lessard \cite{L2011}. It is worth noting that $f_{IJK}/f_{IJ}$ represents the conditional probability for a third sampled individual to be identical by descent to one of two previous sampled individuals, given that those two individuals are identical by descent.

The effect of the population-scaled moments of the viability coefficients in the infinitesimal mean function is encapsulated in the difference of two terms: 
$$\mu_A-\frac{\sigma_A^2}{2}-\left(\mu_B-\frac{\sigma_B^2}{2}\right)+\sigma^2\left(\frac{1}{2}-x\right),$$
 which represents the effect in the absence of relatedness, and
$$\frac{f_{IJK}}{f_{IJ}}\sigma^2\left(\frac{1}{2}-x\right),$$ 
which measures a correcting effect due to relatedness. 
Similarly, the effect of the population-scaled moments of the viability coefficients on the infinitesimal variance function is the difference between $1+\sigma^2x(1-x)$ in the absence of relatedness and a correcting term for relatedness given b y $f_{IJ}\sigma^2x(1-x)$.

Note that the relatedness level in the population is inversely proportional to the deme-scaled dispersal rate $\nu$. As 
$\nu\rightarrow0$, which corresponds to the highest effect of relatedness in the population, the infinitesimal mean and the variance tend to
\begin{subequations}
\begin{align}
M(x,\mathbf{0})&=x(1-x)\left[\mu_A-\frac{\sigma_A^2}{2}-\left(\mu_B-\frac{\sigma_B^2}{2}\right)\right],\\
V(x,\mathbf{0})&=x(1-x).
\end{align}
\end{subequations}
The corresponding diffusion process is similar to the standard Wright-Fisher diffusion observed in unstructured haploid population with deterministic viability coefficients. However, in this scenario, the population-scaled arithmetic means $\mu_A$ and $\mu_B$ are replaced by the population-scaled geometric mean $\mu_A-\sigma^2_A/2$ and $\mu_B-\sigma^2_B/2$, respectively. As a result, selection will favour the fixation of the type associated with the highest population-scaled geometric mean viability coefficient, rendering the population-scaled covariance $\sigma_{AB}$ irrelevant.

It is worth noting that as $\nu$ tends towards infinity, thus removing the influence of structure and migration, the infinitesimal mean and variance functions $M$ and $V$ become identical to those derived by Gillespie \cite{G1973} for two types $A$ and $B$ with fitnesses $1+s_A$ and $1+s_B$, respectively, in an unstructured population, with the exception of the inclusion of $x(1-x)$ in the infinitesimal variance. 

As a final remark, when the population-scaled means, variances, and covariance are relatively small, our calculations indicate that an increase in the population-scaled covariance between $s_A$ and $s_B$ will decrease the fixation probabilities $F_A$ and $F_B$ for both types $A$ and $B$ introduced as single mutants. Note also that an increase in the deme-scales dispersal rate will result in higher fixation probabilities for both types.

\section*{Acknowledgments}
S. Lessard is supported by the Natural Sciences and Engineering Research Council of Canada, grant no. 8833. 

\section{Appendix A}

Owing to Taylor's theorem, there exists $\xi_i$ between $0$ and $\tilde{s}_i$ such that
\begin{equation}\label{aA-eq1}
\frac{1}{(1+\tilde{s}_i)^n}=1-n\tilde{s}_i+\frac{n(n+1)}{2}\tilde{s}_i^2-\frac{n(n+1)(n+2)\tilde{s}_i^3}{3!(1+\xi_i)^{n+3}}.
\end{equation}
Therefore, we have
\begin{align}\label{aA-eq2}
E_{\mathbf{z}}\Big[\tilde{\tilde{x}}_i^n\Big]&=\tilde{x}_i^nE_{\mathbf{z}}\left[\frac{(1+s_A)^n}{(1+\tilde{s}_i)^n}\right]\nonumber\\
&=\tilde{x}_i^nE_{\mathbf{z}}\left[(1+s_A)^n\left(1-n\tilde{s}_i+\frac{n(n+1)}{2}\tilde{s}_i^2-\frac{n(n+1)(n+2)\tilde{s}_i^3}{3!(1+\xi_i)^{n+3}}\right)\right]\nonumber\\
&=\tilde{x}_i^nE_{\mathbf{z}}\left[\left(1+ns_A+\frac{n(n-1)}{2}s_A^2\right)\left(1-n\tilde{s}_i+\frac{n(n+1)}{2}\tilde{s}_i^2\right)\right]+o(D^{-1})\nonumber\\
&=\tilde{x}_i^n+\tilde{x}_i^nE_{\mathbf{z}}\left[n(s_A-\tilde{s}_i)+\frac{n(n+1)}{2}\tilde{s}_i^2+\frac{n(n-1)}{2}s_A^2-n^2s_A\tilde{s}_i\right]+o(D^{-1})\nonumber\\
&=\tilde{x}_i^n+\frac{(1-\tilde{x}_i)\tilde{x}_i^n}{ND}\Big[
n(\mu_A-\mu_B)
+\left(
\frac{n(n-1)}{2}-\frac{n(n+1)}{2}\tilde{x}_i
\right)(\sigma_A^2-\sigma_{AB})\nonumber\\
&\quad\quad\quad\quad\quad\quad\quad\quad\quad\quad\quad+\frac{n(n+1)}{2}(1-\tilde{x}_i)(\sigma_B^2-\sigma_{AB})
\Big]+o(D^{-1}),
\end{align}
for any integer $n\geq1$, from which we obtain
\begin{align}\label{aA-eq3}
E_{\mathbf{z}}\Big[P_{ij}(\mathbf{z})\Big]&=E_{\mathbf{z}}\Big[\binom{N}{j}\left(\tilde{\tilde{x}}_i\right)^j\left(1-\tilde{\tilde{x}}_i\right)^{N-j}\Big]\nonumber\\
&=\binom{N}{j}\sum_{l=0}^{N-j}\binom{N-j}{l}(-1)^lE_{\mathbf{z}}\Big[
\tilde{\tilde{x}}_i^{j+l}\Big]\nonumber\\
&=\binom{N}{j}\sum_{l=0}^{N-j}\binom{N-j}{l}(-1)^l\tilde{x}_i^{j+l}+o(1)\nonumber\\
&=\binom{N}{j}\left(\tilde{x}_i\right)^j\left(1-\tilde{x}_i\right)^{N-j}+o(1).
\end{align}
 Similarly, we have
\begin{align}\label{aA-eq4}
E_{\mathbf{z}}\Big[\tilde{\tilde{x}}_i\tilde{\tilde{x}}_j\Big]&=\tilde{x}_i\tilde{x}_jE_{\mathbf{z}}\left[\frac{(1+s_A)^2}{(1+\tilde{s}_i)(1+\tilde{s}_j)}\right]\nonumber\\
&=\tilde{x}_i\tilde{x}_jE_{\mathbf{z}}\Big[(1+s_A)^2\left(1-\tilde{s}_i+\tilde{s}_i^2\right)\left(1-\tilde{s}_j+\tilde{s}_j^2\right)\Big]+o(D^{-1})\nonumber\\
&=\tilde{x}_i\tilde{x}_jE_{\mathbf{z}}\Big[1+2s_A-\tilde{s}_i-\tilde{s}_j+s_A^2+\tilde{s}_i^2+\tilde{s}_j^2+\tilde{s}_i\tilde{s}_j-2s_A\tilde{s}_i-2s_A\tilde{s}_j\Big]+o(D^{-1})\nonumber\\
&=\tilde{x}_i\tilde{x}_j+\frac{\tilde{x}_i\tilde{x}_j}{ND}\Bigg[
\Big(2-\tilde{x}_i-\tilde{x}_j\Big)(\mu_A-\mu_B)
+\Big(
\tilde{x}_i^2+\tilde{x}_j^2+\tilde{x}_i\tilde{x}_j+1-2\tilde{x}_i-2\tilde{x}_j
\Big)(\sigma_A^2-\sigma_{AB})\nonumber\\
&\quad\quad\quad\quad\quad\quad+\Big(
\tilde{x}_i^2+\tilde{x}_j^2+\tilde{x}_i\tilde{x}_j+3-3\tilde{x}_i-3\tilde{x}_j
\Big)(\sigma_B^2-\sigma_{AB})\Bigg]+o(D^{-1}).
\end{align}
Combining Eq. (\ref{aA-eq2}) for $n=1$ and Eq. (\ref{aA-eq4}), we get
\begin{align}\label{aA-eq5}
Cov_{\mathbf{z}}\Big(\tilde{\tilde{x}}_i,\tilde{\tilde{x}}_j\Big)&=E_{\mathbf{z}}\Big[\tilde{\tilde{x}}_i\tilde{\tilde{x}}_j\Big]-E_{\mathbf{z}}\Big[\tilde{\tilde{x}}_i\Big]E_{\mathbf{z}}\Big[\tilde{\tilde{x}}_j\Big]\nonumber\\
&=\frac{1}{ND}\tilde{x}_i\tilde{x}_j(1-\tilde{x}_i)(1-\tilde{x}_j)\left(\sigma_A^2+\sigma_B^2-2\sigma_{AB}\right)+o(D^{-1}).
\end{align}

\section{Appendix B}

\subsection{Important identities}
We will show three important identities, namely, 
\begin{subequations}
\begin{align}
\label{secA-eq11}&\sum_{j=0}^{N}\tilde{x}_jv_j=x,\\
\label{secA-eq12}&\sum_{j=0}^{N}\tilde{x}_j^2v_j(x)=(1-f)x\left(x-\frac{1}{N}+\frac{1}{m(2-m)N}\right),\\
&\sum_{j=0}^{N}\tilde{x}_j^3v_j(x)=\frac{N^2m^3x^3+(1-m)x\Big[(1-m)^2+3Nm(1-m)x+3N^2m^2x^2\Big]}{N^2-(N-1)(N-2)(1-m)^3}\nonumber\\
\label{secA-eq13}&\quad\quad\quad\quad\quad\quad+\frac{3(N-1)(1-m+mxN)\Big[1+m(2-m)(Nx-1)\Big]xf}{N^2-(N-1)(N-2)(1-m)^3},
\end{align}
\end{subequations}
where 
\begin{equation}
f=\frac{(1-m)^2}{(1-m)^2+mN(2-m)}.
\end{equation}
First, note that 
$\mathbf{v}(x)$ is the solution of the linear system of equations
\begin{equation}\label{secA-eq2}
v_j(x)=\sum_{i=0}^{N}v_i(x)P_{ij}^{*}(x),
\end{equation}
for $j=0,1,\ldots,N$, where $(P_{ij}^*(x))_{j=0}^{N}$ is the probability distribution of a binomial random variable with parameters $N$ and $\tilde{x}_i=ai+b$, for $i=0,1,\ldots,N$,
where $a=(1-m)/N$ and $b=mx$.
Therefore, we have the identities
\begin{subequations}\label{secA-eq3}
\begin{align}
&\sum_{j=0}^{N}jP_{ij}^*(x)=N\tilde{x}_i,\\
&\sum_{j=0}^{N}j^2P_{ij}^*(x)=N\tilde{x}_i+N(N-1)\tilde{x}_i^2,\\
&\sum_{j=0}^{N}j^3P_{ij}^*(x)=N\tilde{x}_i+3N(N-1)\tilde{x}_i^2+N(N-1)(N-2)\tilde{x}_i^3.
\end{align}
\end{subequations}
Using these identities, we obtain the following system of equations:
\begin{subequations}
\begin{align}\label{secA-eq4}
\sum_{j=0}^{N}\tilde{x}_jv_j(x)
&=\sum_{i=0}^{N}v_i(x)\sum_{j=0}^{N}\tilde{x}_jP_{ij}^*(x)
=aN\sum_{i=0}^{N}\tilde{x}_iv_i(x)+b,\\
\label{secA-eq5}
\sum_{j=0}^{N}\tilde{x}_j^2v_j(x)
&=\sum_{i=0}^{N}v_i(x)\sum_{j=0}^{N}\tilde{x}_j^2P_{ij}^*(x)\nonumber\\
&=\sum_{i=0}^{N}v_i(x)\Big[a^2\left( 
N\tilde{x}_i+N(N-1)\tilde{x}_i^2
\right)
+
2abN\tilde{x}_i
+b^2
\Big]
\nonumber\\
&=b^2
+
aN(a+2b)\sum_{j=0}^{N}\tilde{x}_jv_j(x)
+
a^2N(N-1)\sum_{j=0}^{N}\tilde{x}_j^2v_j(x),\\
\label{secA-eq6}
\sum_{j=0}^{N}\tilde{x}_j^3v_j(x)
&=\sum_{i=0}^{N}v_i(x)\sum_{j=0}^{N}\tilde{x}_j^3P_{ij}^*(x)\nonumber\\
&=\sum_{i=0}^{N}v_i(x)\Big[
a^3\left(N\tilde{x}_i+3N(N-1)\tilde{x}_i^2+N(N-1)(N-2)\tilde{x}_i^3\right)\nonumber\\
&\quad+
3a^2b\left(N\tilde{x}_i+N(N-1)\tilde{x}_i^2\right)
+
3ab^2N\tilde{x}_i
+b^3
\Big]\nonumber\\
&=b^3+
aN\left( a^2+ 3ab+3b^2\right)\sum_{j=0}^{N}\tilde{x}_jv_j(x)
 + 3a^2 N(N-1)(a+b)\sum_{j=0}^{N}\tilde{x}_j^2v_j(x)\nonumber\\
 &\quad
  +a^3N(N-1)(N-2)\sum_{j=0}^{N}\tilde{x}_j^3v_j(x).
\end{align}
\end{subequations}
Solving this linear system of equations 
leads to the identities in Eqs. (\ref{secA-eq11})-(\ref{secA-eq13}).

\subsection{Infinitesimal mean and variance}
Making use of the identities in Eqs. (\ref{secA-eq11})-(\ref{secA-eq13}), we obtain
\begin{subequations}\label{secA-eq9}
\begin{align}
&\sum_{j=0}^Nv_j(x)\tilde{x}_j(1-\tilde{x}_j)=(1-f)x(1-x),\\
&\sum_{j=0}^Nv_j(x)\tilde{x}_j^2(1-\tilde{x}_j)=
\frac{(1-f)x(1-x)}{N^2 - (1 - m)^3 (N - 1) (N - 2)
}\\
&\quad\times\left[
\frac{(1-m)^2N(3-m(3-m)(1+x))}{2-m}+mN^2x(3-3m+m^2)-(1-m)^2
\right],\nonumber\\
&\sum_{j=0}^Nv_j(x)\tilde{x}_j(1-\tilde{x}_j)^2=
\frac{(1-f)x(1-x)}{N^2 - (1 - m)^3 (N - 1) (N - 2)
}\nonumber\\
&\quad\times\left[\frac{(1 - m)^2 N (3 - m (3 - m) (2 - x))}{2-m}+
(3-3m+m^2)(1-x)mN^2-(1-m)^3
\right].
\end{align}
\end{subequations}
Then, the infinitesimal mean and variance of the limiting diffusion process  are given by
\begin{align}
M(x,\mathbf{0})&=\sum_{j=0}^{N}v_j(x)\tilde{x}_j(1-\tilde{x}_j)\Big[\mu_{A}-\mu_{B}+\tilde{x}_j(\sigma_{AB}-\sigma^2_{A})+(1-\tilde{x}_j)\left(\sigma^2_{B}-\sigma_{AB}\right)\Big]\nonumber\\
&=(1-f)x(1-x)(\mu_{A}-\mu_{B})\nonumber\\
&\quad+\frac{(1-f)x(1-x)}{N^2 - (1 - m)^3 (N - 1) (N - 2)
}\Big[
\frac{(1-m)^2N(3-m(3-m)(1+x))}{2-m}\nonumber\\
&\quad\quad\quad\quad\quad\quad\quad\quad\quad\quad+mN^2x(3-3m+m^2)-(1-m)^2
\Big]
(\sigma_{AB}-\sigma^2_{A})\nonumber\\
&\quad+\frac{(1-f)x(1-x)}{N^2 - (1 - m)^3 (N - 1) (N - 2)
}\Big[
\frac{(1-m)^2N(3-m(3-m)(2-x))}{2-m}\nonumber\\
&\quad\quad\quad\quad\quad\quad\quad\quad\quad\quad+mN^2(1-x)(3-3m+m^2)-(1-m)^2
\Big]
(\sigma^2_{B}-\sigma_{AB})
\end{align}
and 
\begin{align}
V(x,\mathbf{0})&=\left(\sum_{i=0}^{N} v_i(x)\tilde{x}_i(1-\tilde{x}_i)\right)^2\left(\sigma_{A}^2+\sigma_B^2-2\sigma_{AB}\right)+\sum_{i=0}^{N}v_i(x)\tilde{x}_i(1-\tilde{x}_i)\nonumber\\
&=(1-f)x(1-x)\Big[1+(1-f)x(1-x)\left(\sigma_{A}^2+\sigma_B^2-2\sigma_{AB}\right)\Big],
\end{align}
respectively.

\bibliographystyle{unsrt}

\end{document}